\documentclass[runningheads]{llncs}
\usepackage[T1]{fontenc}
\usepackage{graphicx}
\usepackage{amsmath}
\usepackage{amssymb}
\usepackage{multirow}
\usepackage{booktabs}
\usepackage{subcaption}
\usepackage{array}

\usepackage{color, colortbl}
\usepackage{tabularx}

\usepackage{xcolor} 
\usepackage[colorlinks=false, 
            linkbordercolor=red, 
            citebordercolor=green, 
            urlbordercolor=cyan, 
            pdfborder={0 0 1} 
           ]{hyperref}

\usepackage{orcidlink}

\newcommand{\blueurl}[1]{{\color{blue}\href{#1}{#1}}}

\renewcommand{\url}[1]{\blueurl{#1}}

\newcolumntype{L}{>{\raggedright\arraybackslash}X}
\newcolumntype{R}{>{\raggedleft\arraybackslash}X}
\newcolumntype{Y}{>{\centering\arraybackslash}X}

\definecolor{HighlightR}{rgb}{0.71, 0.15, 0.11} 
\definecolor{HighlightG}{rgb}{0.15, 0.71, 0.11}
\definecolor{HighlightB}{rgb}{0.0, 0.40, 0.58} 
\definecolor{HighlightY}{rgb}{0.99, 0.87, 0.17}

\newcommand{\set}{\mathbb}

\newcommand{\PatVQA}{\textsc{PatFigVQA}}
\newcommand{\PatCLS}{\textsc{PatFigCLS}}

\newcommand{\subgraphs}[3]{%
    \begin{subfigure}[b]{0.48\textwidth}
        \centering
        \includegraphics[height=5cm]{#1}
        \label{#2}
    \end{subfigure}%
    \hspace{0.02\textwidth}%
}

\newcommand{\subgraphscfmatrix}[3]{%
    \begin{subfigure}[b]{0.49\textwidth}
        \centering
        \includegraphics[height=4.5cm]{#1}
        \label{#2}
    \end{subfigure}%
    \hspace{0.01\textwidth}%
}

\usepackage{array}
\usepackage{adjustbox}

\newcommand{\imageTextRow}[3]{%
    \begin{tabular}{@{}m{0.15\textwidth}m{0.8\textwidth}@{}}
        \adjustbox{valign=t}{\includegraphics[height=#1, keepaspectratio]{#2}} &
        \begin{minipage}[t]{\linewidth}
            \fontsize{6}{8}\selectfont #3
        \end{minipage} \\[1em]
        \midrule
    \end{tabular}
}

\begin{document}
\title{Patent Figure Classification using Large Vision-language Models}
\titlerunning{Patent Figure Classification using LVLMs}

\author{Sushil Awale\inst{1}\orcidlink{0000-0003-2575-0134} \and
Eric M\"{u}ller-Budack\inst{1}\orcidlink{0000-0002-6802-1241} \and
Ralph Ewerth\inst{1,2}\orcidlink{0000-0003-0918-6297}}
\authorrunning{Awale et al.}
\institute{TIB – Leibniz Information Centre for Science and Technology, Hannover, Germany \and
L3S Research Center, Leibniz University, Hannover, Germany\\
\email{\{sushil.awale,eric.mueller,ralph.ewerth\}@tib.eu}}
\maketitle              
\begin{abstract}
Patent figure classification facilitates faceted search in patent retrieval systems, enabling efficient prior art search. Existing approaches have explored patent figure classification for only a single aspect and for aspects with a limited number of concepts.
In recent years, large vision-language models (LVLMs) have shown tremendous performance across numerous computer vision downstream tasks, however, they remain unexplored for patent figure classification.
Our work explores the efficacy of LVLMs in patent figure visual question answering (VQA) and classification, focusing on zero-shot and few-shot learning scenarios. For this purpose, we introduce new datasets, \textsc{\PatVQA{}}~and \textsc{\PatCLS{}}, for fine-tuning and evaluation regarding multiple aspects of patent figures~(i.e., type, projection, patent class, and objects). 
For a computational-effective handling of a large number of classes using LVLM, we propose a novel tournament-style classification strategy that leverages a series of multiple-choice questions. 
Experimental results and comparisons of multiple classification approaches based on LVLMs and Convolutional Neural Networks (CNNs) in few-shot settings show the feasibility of the proposed approaches. 

\end{abstract}
\section{Introduction}
\label{sec:intro}

Patent figures, mostly binary or grayscale, help illustrate innovations using varied figure types e.g., technical drawings, graphs~\cite{CLEFIP2011} and projections e.g., cross-sectional, elevational~\cite{VIEWPOINT_TEXT}. Patent figures and technical illustrations specifically serve the purpose of communicating information more effectively than text alone~\cite{Carney2002}. Despite their importance, prior works in patent domain have largely focused on text modality~\cite{KRESTEL2021102035} for tasks such as patent classification~\cite{PatentNetMC,PatCLS}, patent retrieval~\cite{SEARCHFORMER}, and patent summarization~\cite{patSum}. However, incorporating visual modality alongside text can enhance the performance of these tasks~\cite{CsurkaRJ11,pustu2021multimodal,multimodalRetrieval}. For instance, figure classification in patent retrieval systems, would enable faceted search, aiding patent examiners find relevant patents more efficiently~\cite{KRESTEL2021102035}. Considering the pivotal role of figures in patents and their potential benefits in downstream tasks, developing a robust classifier for categorizing figures into various aspects is essential.
\begin{figure}[!tb]
    \centering
    \includegraphics[width=1.0\textwidth]{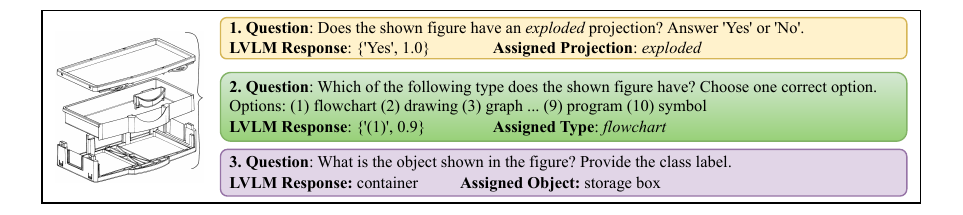}
    \caption{\small A figure from patent \textsc{USD534354S1} showing a \textit{drawing} of a \textit{modular tool storage drawer} in \textit{cross-sectional projection}, and three different questions asking about various aspects of the figure: 1) Binary question asking about projection 2) Multiple-choice question asking about figure type, and 3) Open-ended question asking about object depicted. The figure also shows the response~(and token probability) generated from an LVLM and the corresponding concept assigned to the figure.}
    \label{fig:example}
\end{figure}

Patent figure classification differs significantly from natural image classification. Patent figures share common traits with scientific diagrams such as plain white background, abstractness, sparseness, etc.~\cite{KucerOC022}, however, they have distinct features such as specialized projections~(e.g., exploded, cross-sectional, and partial) or leading lines that connect reference numbers to specific components\footnote{\url{https://www.uspto.gov/web/offices/pac/mpep/s1825.html}}. Figure~\ref{fig:example} shows an example of a patent figure in~\textit{exploded projection}. 
Prior works have studied classification of figures into type~\cite{GhauriME23,Kravets2017/12}, projection~\cite{GhauriME23}, IPC~(International Patent Classification)\footnote{\url{https://www.wipo.int/classifications/ipc/en/}} class~\cite{Kravets2017/12}, and object~\cite{OBJCLS,VROCHIDIS2012292,VROCHIDIS201094} using traditional and deep learning methods.
In these works, patent figures were classified into a handful of categories. For example, Ghauri et al.~\cite{GhauriME23} classify figures up to seven projection categories, and Kravets et al.~\cite{Kravets2017/12} classify figures into three IPC classes. On the other hand, large vision-language models~(LVLMs) such as~\textsc{InstructBLIP}~\cite{InstructBLIP},~\textsc{LLaVA}~\cite{LLaVA} in patent figure classification remains unexplored, despite their success in analyzing scientific graphs and charts~\cite{LEAFQA,ChartQA}.

In this paper, we investigate to what extent LVLMs, pretrained on natural images, can classify patent figures by type, projection, object, and USPC~(United States Patent Classification)\footnote{\url{https://www.uspto.gov/patents/search/classification-standards-and-development}} class. 
Our main contributions are as follows: (1)~We introduce a novel visual question answering~(\textit{VQA}) dataset called~\textsc{\PatVQA{}} suitable to fine-tune and evaluate LVLMs in a few-shot learning scenario to close the domain gap of pre-trained LVLMs for patents. (2)~Additionally, we propose a novel LVLM-based tournament-style classification approach, which leverages a series of multiple-choice questions to efficiently perform patent figure classification for a large number of classes. (3)~We perform a comparative study between this approach and CNN-based classifier for patent figure classification on a new dataset called~\textsc{\PatCLS{}}. Experimental results show promising results of the proposed classifier compared to binary and open-ended classification~(see Figure~\ref{fig:example}). Furthermore, it outperforms the CNN-based classifiers for two out of four classification aspects. 
The dataset and source code are publicly available at~\url{https://github.com/TIBHannover/patent-figure-classification}.

The remainder of the paper is organized as follows. Section~\ref{sec:rw} reviews related work on patent figure classification, and use of LVLMs for figure analysis. In Section~\ref{sec:methodology}, we discuss on prompts for figure classification, formulation of a figure VQA dataset~\PatVQA, and different LVLM-based figure classification approaches. Section~\ref{sec:experiments} discusses the implementation details of our experiments and analyse the results of our VQA and classification experiments. Finally, Section~\ref{sec:conclusion_and_future_work} concludes the paper and outlines potential directions for future work.

\section{Related Work}
\label{sec:rw}

Recent research has explored figure classification and chart analysis in scientific and technical documents~\cite{FigureCLSSurvey,ChartSurvey,DocFigure}, including patents, using both both traditional and deep learning methods~\cite{KRESTEL2021102035}. In recent years, LVLMs have shown substantial performance in multimodal tasks~\cite{VisualGPT,InstructBLIP,Hu_2022_CVPR,BLIP2,LLaVA}, including scientific diagram and chart analysis~\cite{li2023scigraphqa,ChartQA,PlotQA}. This section reviews related work on patent figure classification~(Section \ref{sec:rw_fig_cls}) and LVLMs for scientific and technical figure analysis~(Section~\ref{sec:rw_lvlms}). 

\subsection{Patent Figure Classification}
\label{sec:rw_fig_cls}

The~\textbf{CLEF-IP 2011} (Cross-Language Evaluation Forum - Intellectual Property track) dataset contains manually labeled patent figures categorized into nine different figure types~\cite{CLEFIP2011}. The dataset enabled numerous works on classification of patent figures by type. Early approaches used Fisher vectors~\cite{FisherVectors} with linear classifiers~\cite{CsurkaRJ11}. Recent methods use deep learning techniques such as CNN~\cite{DETC2020,Kravets2017/12}. For instance, Jiang~\cite{DETC2020} classify patent figures by type and IPC class using Dual VGG19~(Visual Geometry Group) network~\cite{VGG19}.

Ghauri et al.~\cite{GhauriME23} extend the~\textbf{CLEF-IP 2011} dataset by adding a new figure type~\textsc{block\_or\_circuit}, and compare different CNN-based models with CLIP~(Contrastive Language-Image Pretraining)~\cite{RadfordKHRGASAM21} on this enhanced dataset. Ghauri et al.~\cite{GhauriME23} also created the~\textbf{USPTO-PIP}~(United States Patent and Trademark Office-Patent Image Perspective) for two-level projection classification and classify patent figures by projections. While patent figure classification by object, CPC~(Cooperative Patent Classification)\footnote{\url{https://www.cooperativepatentclassification.org/about}}, and~USPC class remain understudied, some research has focused on learning patent figure representations for concept-based retrieval~\cite{VROCHIDIS2012292,VROCHIDIS201094}. 

\subsection{Large Vision-language Models for Technical Diagrams}
\label{sec:rw_lvlms}

LVLMs have demonstrated strong performance across a wide array of computer vision tasks such as image captioning~\cite{VisualGPT,Hu_2022_CVPR}, scene understanding~\cite{BLIP2,LLaVA}, visual question answering~\cite{InstructBLIP,BLIP2,LLaVA}, etc. In recent works, Ging et al.~\cite{gingbravo2024ovqa} perform a comparative study of different LVLMs for classification of natural images. 

In the context of scientific domains, LVLMs have been studied for analysing charts and graphs~\cite{ChartQA,PlotQA}. They have shown robust performance for graph datasets such as \textsc{ScieneQA+}~\cite{li2023scigraphqa}, \textsc{ChartQA}~\cite{ChartQA}, and \textsc{PlotQA}~\cite{PlotQA}. These tasks are more complex than classification as they require visual comprehension and reasoning. However, LVLMs remain unexplored for patent figures.


\section{Patent Figure Classification using LVLMs}
\label{sec:methodology}

\begin{figure*}[t]
    \centering
    \includegraphics[width=1.0\linewidth]{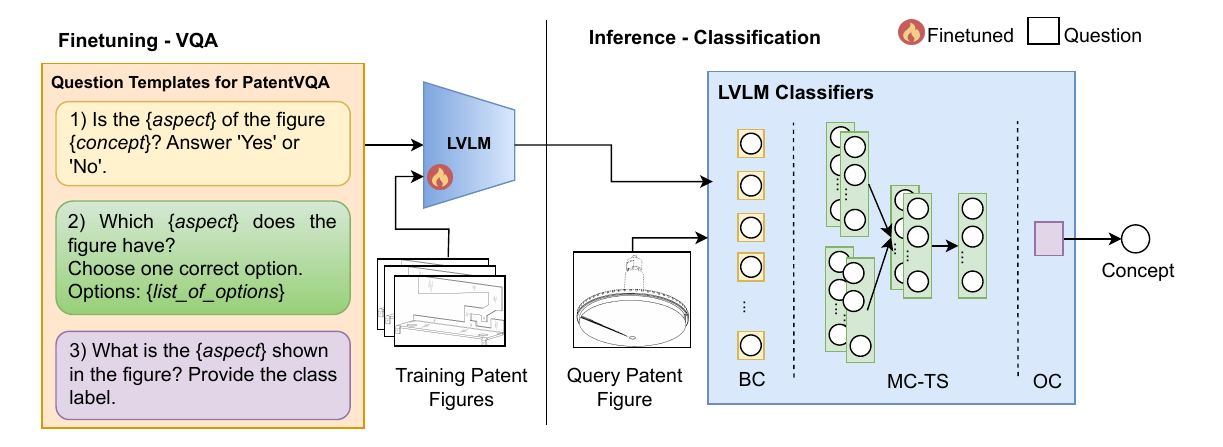}
    \caption{\small Workflow of patent figure classification using different LVLM-based classification approaches. On the left, question templates for different question types used to create~\PatVQA~dataset are shown. On the right, three different approaches to figure classification using a fine-tuned LVLM is shown, which include Binary Classification [BC], Multiple-choice Classification - Tournament-style Strategy [MC-TS], and Open-ended Classification [OC].}
    \label{fig:cls_fig}
\end{figure*}

In this section, we present our approach for patent figure classification using LVLMs. 
In Section~\ref{sec:prompts}, we discuss the different types of prompt for figure classification. Section~\ref{sec:vqa_dataset} describes the methodology to create a patent figure VQA dataset used to fine-tune a pre-trained LVLM for patent classification. Finally, Section~\ref{sec:cls} details three distinct approaches of figure classification that leverage LVLM's VQA capability. Figure~\ref{fig:cls_fig} shows an overview of the workflow of patent figure classification using LVLM.

\subsection{Prompt Type for Patent Figure Classification } 
\label{sec:prompts}

For figure classification, the goal is to classify a patent figure~$f$ to a concept~$c$ from a set of concepts~$\set{C}_a$ of an aspect~$a\in\set{A}$.
We consider classifying patent figures on multiple aspects $a \in \set{A} = \{\text{\textsc{Type},~\textsc{Projection},~\textsc{Object}, \textsc{USPC}}\}$. \textsc{TYPE} denotes the form or representation of the visual element depicted in a figure~\cite{CLEFIP2011}.~\textsc{PROJECTION} in a figure is the viewing angle or perspective in which a diagram or a drawing is portrayed~\cite{Carlbom1978}.~\textsc{OBJECT} illustrates the concept depicted in the figure. And,~\textsc{USPC} denote the class in the hierarchical \textit{United States Patent Classification Scheme}.

To enable LVLMs for figure classification, along with a figure, a text prompt is required as input.
For this purpose, we consider the following three types of distinct natural questions as prompts~$q$ that elicits a response~$y$ from a model~$\psi$ about the conceptual content represented in a figure~$f$:
\begin{enumerate}
    \item \textsc{Binary}~($q_b$): questions that require a binary response~$y \in \{yes, no\}$;
    \item \textsc{Multiple-choice}~($q_m$): questions that ask for the most likely concept from a (sub)set of predefined concepts~$\set{C}_a^{'} \subset \set{C}_a$ as response~$y \in \set{C}_a^{'}$; 
    \item \textsc{Open-ended}~($q_o$): questions that allow for a free-form response~$y$.
\end{enumerate}
Examples of each question type are shown in Figure~\ref{fig:example}. Parallel to our work, these prompting strategies have been studied for passage re-ranking using LLMs~\cite{setwise}.

For figure classification, the response~$y = \psi(f, q)$ from the model~$\psi$ maps to a concept~$c$ from a concept set~$\set{C}$ for a figure~$f$ by a classification function~$\phi(f, y)~\mapsto~c\in\set{C}$.
With these three distinct question types and a corresponding classification function~$\phi$~(detailed in Section~\ref{sec:cls}), we evaluate a pre-trained LVLM on patent figure classification task in a zero-shot setting. 

\subsection{\textbf{\textsc{\PatCLS}~and~\textsc{\PatVQA}~datasets}}
\label{sec:vqa_dataset}
LVLMs use image encoders such as ViT-g/14~\cite{Eva} in~\textsc{InstructBLIP}~\cite{InstructBLIP}, which are largely pre-trained on natural images, and are not adapted for patent figures. Hence, to enhance the LVLMs ability to acclimate to unique features of patent figures, we suggest to fine-tune LVLMs to the task of patent figures VQA. To this end, first we construct a patent figure classification dataset~$-$~\PatCLS~and then transform it to a patent figure VQA dataset~$-$~\PatVQA. 
For this purpose, we adapt two existing classification datasets 
\textbf{(1) Extended CLEF-IP 2011}~\cite{GhauriME23}, which contains $35,926$ utility patent figures classified into $10$ different figure \textsc{Types}, and
\textbf{(2) DeepPatent2}~\cite{Ajayi2023}, which comprises of~$2,785,762$ segmented industrial design patent figures covering~$22,394$ unique~\textsc{Projections},~$132,890$ unique \textsc{Objects}, and~$33$ \textsc{USPC} classes.

The~\textsc{Projection} and the~\textsc{Object} concepts in the~\textbf{DeepPatent2} dataset were extracted automatically from figure references. Therefore, the concepts are not normalized, and hence not suitable for classification purposes directly. To mitigate this problem, in case of~\textsc{Projections}, we map the extracted concepts to a pre-defined projection classification schema, which is based on Carlbom et al.~\cite{Carlbom1978}. We extend the schema by grouping~\textit{multiview} projections into four specialized views:~\textit{elevation} and~\textit{plan} based on Radford's architectural drawing~\cite{radford1912architectural}, and~\textit{sectional} and~\textit{detail}, which are common projections in design patent drawings\footnote{\url{https://www.uspto.gov/patents/basics/apply/design-patent\#rules}}. We map the~\textsc{Projection} concepts to the schema using rule-based keyword matching. For example, concepts with keywords~\textit{left perspective} and~\textit{cross-sectional} are mapped to~\textit{perspective} and~\textit{sectional} projections, respectively. 

For~\textsc{Object}, we begin by embedding the set of concepts~$\set{C}$ and clustering the resulting concept embeddings. Next, we map each~\textsc{Object} concept to a cluster, using the concept closest to the cluster centroid as the representative concept. Additionally, in the case of~\textsc{USPC}, we make a modification by excluding the “Miscellaneous” concept due to its lack of specificity. To finalize the~\textit{train} set for our classification dataset, we apply a filtering criterion: we retain only those concepts that have at least $150$ figures each. This ensures that every concept across all aspects has a minimum of $150$ figures per concept and $50$ figures per concept for each question type, making the dataset well-suited for studying few-shot patent figure classification. The resulting~\textit{train} set comprises of~$10$~\textsc{Type}, ~$7$~\textsc{Projection}, $32$~\textsc{USPC}, and $1447$~\textsc{Object} concepts.

For the~\textit{validation} and~\textit{test} sets for~\textsc{Projection}, \textsc{Object}, and \textsc{USPC}, we initially sample at least~$1$ figure per concept and continue further until~$1000$ figures are sampled for each test sets. In case of~\textsc{Object}, we have $1,447$ test samples, one for each concept. For~\textsc{Type}, we use the same~\textit{validation} and~\textit{test} splits as Ghauri et al.~\cite{GhauriME23}. 
Table~\ref{tab:stats} shows the distribution of the dataset across each aspect. We refer to this classification dataset as~\PatCLS~dataset.
\begin{table}[t]
  \setlength{\tabcolsep}{8pt}
  \centering
  \caption{Number of~\underline{train}ing,~\underline{valid}ation, and~\underline{test} samples and number of concepts for~\PatCLS{} dataset}
  \label{tab:stats}
  {\small{
      \begin{tabularx}{1.0\linewidth}{lRRRRR}
        \toprule
        & & \multicolumn{3}{c}{Splits} \\
        \cmidrule{3-5}
        \multirow{-2}{*}{Aspect} & \multirow{-2}{*}{Concepts} & Train & Valid & Test \\
        \midrule
        Type & 10 & 1,500 & 180 & 1,040 \\
        Projection & 7 & 1,050 & 1,000 & 1,000 \\
        USPC & 32 & 4,800 & 1,000 & 1,000 \\
        Object & 1,447 & 217,050 & 1,447 & 1,447 \\
        \bottomrule
      \end{tabularx}
  }}
\end{table}

Next, using the same samples from~\PatCLS~dataset, we create~\PatVQA. For this purpose, we manually formulate three distinct types of natural language question templates (shown in Figure~\ref{fig:cls_fig}) for each prompt type discussed in Section~\ref{sec:prompts}, and formulate a valid question by replacing the placeholder in the template with the corresponding concept of the figure sampled from~\PatCLS. 
For~\textsc{Binary} questions, for example, we replace the placeholders~\textit{\{aspect\}} and~\textit{\{concept\}} with the corresponding concept~$c$ and aspect~$a$ of the sampled figure~$f$ in the template:~\textit{Is the \{aspect\} of the figure \{concept\}? Answer ‘Yes’ or ‘No’.}
For~\textsc{Multiple-choice} questions, we create questions with different count of options~$K \in \{5, 10, 20\}$, where the list of options include the correct concept and other concepts sampled from a concept set~$\set{C}$. In case of~\textsc{Objects}, we sample the options from a subset~$\set{C}_c$ of $100$ most similar concepts for concept~$c$. The subset~$\set{C}_c$ is constructed based on cosine similarity score with PatentBERT~\cite{srebrovic2020leveraging} concept embeddings. We enumerate the list of options with Roman numerals enclosed by round brackets~$()$, and replace the placeholder~(\textit{\{list\_of\_options\}}) in the template.
In case of~\textsc{Open-ended} questions, placeholders are not required.
Next, we postpend short instructions like “Answer ‘Yes’ or ‘No’.”, “Choose one option.”, and “Provide a class label.” for corresponding questions which instruct the LVLM to generate a desired response. 
In the~\textit{train} set for~\PatVQA, we ensure a balanced distribution of question types by pairing each binary and open-ended question with one multiple-choice question, which can have 5, 10, or 20 options. In the~\textit{validation} and~\textit{test} sets, we sample from the corresponding sets in~\PatCLS, where for each sample we formulate questions for each question type.

\subsection{Figure Classification Approaches using LVLMs}
\label{sec:cls}
For classification of patent figures, we devise three distinct classification approaches, each based on the different question types discussed in Section~\ref{sec:prompts}. Depending on the question type, the classification function~$\phi$ requires one or more queries to the model~$\psi$.

\subsubsection{Binary Classification \textsc{[BC]}}
\label{sec:bc}
approach classifies a concept~$c^i$ to a figure~$f$ using a series of~\textsc{Binary} questions~$Q_b = \{q_b^i : i \in \mathbb{N}, 1 \leq i \leq |\set{C}|\}$, where each question~$q_b^i$ corresponds to a concept~$c^i$ in a concept set~$\set{C}$.
This approach is comparable to~\textit{Pairwise} strategy in passage re-ranking~\cite{qin-etal-2024-large}.

In~\textsc{BC}, a concept~$c^i$ associated with an affirmative response~$y^i$ from the LVLM~$\psi$ is selected as the concept depicted in a figure~$f$. In instances of multiple positive responses, the response with the highest log-likelihood score, derived from the model's output~\textsc{logits}, is selected.
The~\textsc{BC} approach is computationally expensive and is difficult to scale for large concept sets. It requires~$N = |\set{C}|$ queries for classification of a single figure.

\subsubsection{Open-ended Classification \textsc{[OC]}}
\label{sec:oc}
unlike~\textsc{BC} classification approach, requires a single LVLM query~($N=1$). In this approach, an~\textsc{Open-ended} question~$q_o$ is used to query the LVLM~$\psi$ to generate a response~$y \rightarrow \{text\}$ allowing the LVLM to use tokens from the large vocabulary set seen during the training phase. However, it is quite challenging as the question is open-ended, and the vocabulary set seen during training can be large. Moreover, response text and the ground truth may match semantically but not syntactically, which requires further semantic similarity matching~$sim(y, c_n)$ to map the response~$y$ to a concept~$c$ from the concept set~$\set{C}$ (discussed in Section~\ref{sec:implt_details}).

\subsubsection{Multiple-choice Classification \textsc{[MC]}}
\label{sec:mc}
approach selects a concept~$c_i$ from a list of enumerated concepts in a~\textsc{Multiple-choice} question and assigns it to a figure~$f$. The concepts in the question are drawn from a set of concepts~$\set{C}$. This method is analogous to~\textit{Setwise} strategy employed in passage reranking task~\cite{setwise}. In this approach, a multiple-choice question~$q_m^{\set{C}}$ is constructed, incorporating all concepts from the set~$\set{C}$. However, LLMs are constrained by their context length, and prior studies~\cite{Liu2023LostIT} have demonstrated a decline in performance as the context length increases. Contrasting to the~\textit{Setwise} method, we introduce a new strategy to mitigate the limitation of context length:

\textbf{The Tournament-style Strategy~\textsc{[MC-TS]}} partitions the concept set~$\set{C}$ into smaller subsets and query the LVLM with a series of~\textsc{Multiple-choice} questions, each covering a subset of concepts. Here, we partition the concept set~$\set{C}$ to~$Z$ number of subsets i.e. $\set{C} = \bigcup_{j=1}^Z \set{C}^j, \quad \text{where } |\set{C}^j| \leq k$ and $k$ is the maximum size of the subset. We then construct a number of~\textsc{multiple-choice} questions~$Q_m = \{q_m^j : j \in \mathbb{N}, 1 \leq j \leq Z\}$ one for each subset. We then query the LVLM~$\psi$ for each subset
and the most likely concept~$c^j$ from each subset enters a tournament-style process, where the concepts form a new set for a new round of queries.
In each subsequent round~$r$, a new~\textsc{Multiple-choice} question~$q_m$ is posed to the LVLM~$\psi$, incorporating the most likely concepts from the previous rounds. The tournament-style strategy allows for a hierarchical decision-making process, where each round builds upon the results of previous one. The number of rounds~$R$ required for a single figure classification is contingent on the maximum size~$k$ of a subset and the total size of the superset i.e.~$|\set{C}|$. It is computed as~$R = \lceil \log_k{|\set{C}|}\rceil$
and the number of required queries is~$N=\sum_{r=1}^R \left\lceil \frac{|\set{C}|}{k^r} \right\rceil$ The process continues until a single concept remains, which is then assigned as the final classification for the figure~$f$.

\section{Experimental Setup and Results}
\label{sec:experiments}

Our objective in this work is two-fold. First, we study patent figure VQA in zero-shot and few-shot scenarios. Second, we compare different LVLM-based approaches for patent figure classification. With this goal, Section~\ref{sec:implt_details} details the implementation settings of the experiments. In Section~\ref{sec:fine_tuning} and~\ref{sec:cls_evaluation}, we describe the fine-tuning of the LVLM and classification experiments along with their results respectively.

\subsection{Implementation Details}
\label{sec:implt_details}

We run all our experiments for LVLM-based approaches using~\textsc{InstructBLIP}~\cite{InstructBLIP} as the LVLM with~\textsc{FlanT5-XL}~\cite{FlanT5} as the model's LLM backbone. \textsc{InstructBLIP} has proven to be very capable in numerous VQA tasks~\cite{gingbravo2024ovqa}. Moreover, the LLM bakbone~\textsc{FlanT5} was fine-tuned on multiple Natural Language Processing benchmarks containing multiple-choice question answering and classification datasets~\cite{InstructBLIP}. Ging et al.~\cite{gingbravo2024ovqa} also show superior performance of~\textsc{InstructBLIP} on real-world image classification task compared to other LVLMs. Additionally, the vision encoder and the LLM are frozen during fine-tuning, and only a projection layer named Q-former with $188$ million parameters is updated making it efficient to fine-tune. All this pre-training and features, makes~\textsc{InstructBLIP} a suitable choice of LVLM for figure classification task. 

For~\textsc{Open-ended} classification and VQA tasks, we implement the semantic similarity matching function~(discussed in Section~\ref{sec:oc}) using cosine similarity on PatentBERT embeddings~\cite{srebrovic2020leveraging}.

For CNN-based classifiers, following Ghauri et al.~\cite{GhauriME23}, we fine-tune the last layers of~\textsc{ResNet50}~\cite{ResNet50} and~\textsc{ResNext101}~\cite{ResNext101} for $30$ epochs with batch sizes of $512$ and $256$ respectively. We use SGD optimizer with a momentum of $0.9$ and a learning rate of $1e - 3$ with cosine annealing with warm restarts. For fine-tuning of~\textsc{InstructBLIP}, we use the official LAVIS library~\cite{lavis}, and train for $10$ epochs with a batch size of $128$ and AdamW optimizer~\cite{loshchilov2018decoupled} with a weight decay of $0.05$. We apply cosine learning rate scheduler with a learning rate of $1e - 5$. For both classification and VQA task, we select the best model based on accuracy on the validation data. All experiments are conducted using a single NVIDIA H100 GPU with 80GB VRAM.

\subsection{Results for Visual Question Answering}
\label{sec:fine_tuning}
In these experiments, we perform a comparative study on the performance of~\textsc{InstructBLIP} between zero-shot and few-shot setting for figure VQA task. For the few-shot setting, we fine-tune~\textsc{InstructBLIP} for each individual aspects on the~\PatVQA~dataset with increasing training samples~($9 - 150$) per concept. Here, the training samples are balanced across each question type. For the VQA task, we report the accuracy on exact string matching metric.
\begin{figure*}[t]
    \centering
    \begin{subfigure}{\textwidth}
        \centering
        \subgraphs{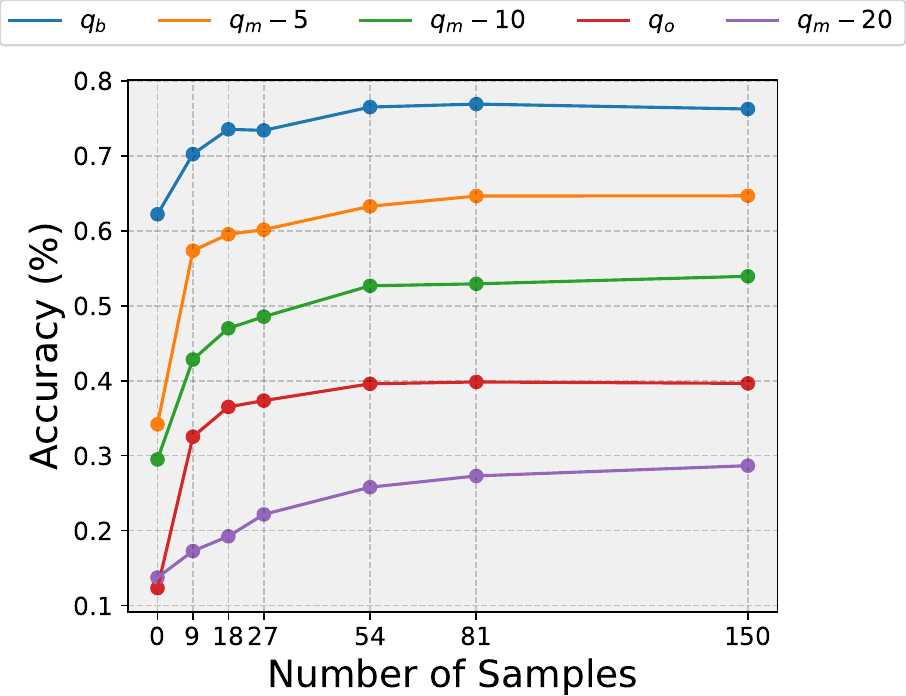}{}{fig:by_task}
        \subgraphs{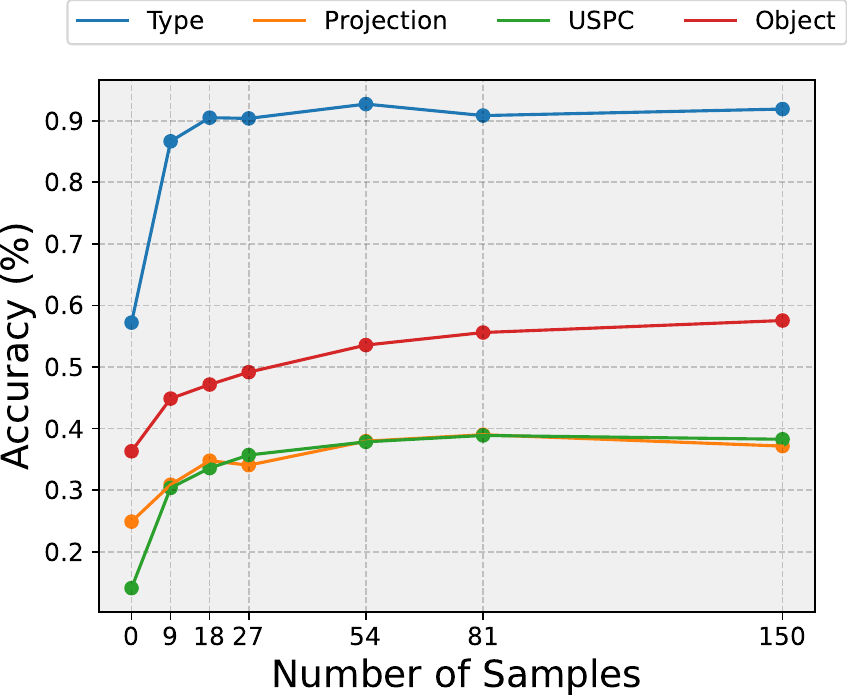}{}{fig:by_aspect}
    \end{subfigure}
    \caption{Average exact string matching accuracy of~\textsc{InstructBLIP} (left) across different question types, and (right) across different aspects on the~\PatVQA~dataset with increasing number of samples per concept.}
    \label{fig:fs_acc}
\end{figure*}
As shown in Figure~\ref{fig:fs_acc} the performance of~\textsc{InstructBLIP} improves as the number of samples increases across each aspect and across each question type. However, after approximately~$80$ training samples per concept the performance starts to saturate. For different question types, the~\textsc{Binary} question, which is the most simple question among them, demonstrates the best performance. In contrast, the performance of~\textsc{Multiple-choice} questions degrades with increasing number of options in the question, which confirms prior studies~\cite{Liu2023LostIT} indicating a drop in performance with longer context. Specifically, for~\textsc{Multiple-choice} questions with~$20$ options, the accuracy is even lower compared to that of~\textsc{Open-ended} questions.

In case of performance by aspects,~\textsc{Type} exhibits the highest accuracy with only~$10$ distinct concepts. While a lower number of concepts might signal a relatively easier task, this is not reflected for the~\textsc{Projection} and~\textsc{USPC} aspects. One reason might be that the difference between different~\textsc{Projections} can be very subtle, which the model fails to capture as it may focus more on the conceptual depiction in the figure. The~\textsc{USPC} concepts, as well, represent the top class in the patent classification hierarchy and have broad definitions, resulting in less precise classification boundaries. Conversely, the pre-training of LVLM is significant, as~\textsc{Objects} outperform both~\textsc{USPC} and~\textsc{Projection} aspects, which is a general aspect observed in natural images.

\subsection{Results for Patent Figure Classification}
\label{sec:cls_evaluation}

In our experiments, we assess the performance of our proposed figure classification method, \textsc{MC-TS}, against other LVLM-based classification approaches outlined in Section~\ref{sec:cls} and CNN-based supervised classifier using the \PatCLS~dataset. In the~\textsc{MC-TS} approach, we experiment with $5$, $10$, and $20$ options~(\textsc{MC-TS~(5)},~\textsc{MC-TS~(10)},~\textsc{MC-TS~(20)}). We use Top-1 accuracy as our primary evaluation metric. For the \textsc{USPC} and \textsc{Object} aspects, we introduce an additional metric: the percentage of samples where an LLM deems the selected concept and the reference concept as semantically equivalent (\textsc{SemEq}). This semantic equivalency metric is adapted from the LAVE metric~\cite{LAVE}, originally developed for VQA evaluation. Following~\textsc{LAVE}, we also use \textsc{FlanT5-XXL}~\cite{FlanT5} as the LLM evaluator. To validate the \textsc{SemEq} metric, we calculated Cohen's Kappa ($\kappa=0.59$) and Inter-Annotator Agreement (IAA = $77$\%) between the LLM and a human annotator, based on a random sample of $200$ test results for the \textsc{Object} aspect using the \textsc{MC-TS~(5)} approach.
\begin{table}
    \centering
    \caption{Top-1 accuracy and~\textsc{SemEq} of different LVLM-based figure classification approaches in zero-shot and few-shot settings~($n$ denotes the maximum number of training samples per class) for Type, Projection (Proj.), USPC, Objects.}
    \label{tab:cls_results}
    \scriptsize
    \setlength{\tabcolsep}{2pt}
    \begin{tabularx}{\linewidth}{l|c|Y|Y|Y|Y|Y|Y}
      \toprule
      \multirow{2}{*}{\textbf{Approach}} & \multirow{2}{*}{\textbf{n}} & \textbf{Type (10)} & \textbf{Proj. (7)} & \multicolumn{2}{c|}{\textbf{USPC (32)}} & \multicolumn{2}{c}{\textbf{Objects (1,447)}} \\
      & & Top-1 & Top-1 & Top-1 & SemEq & Top-1 & SemEq \\
      \toprule
      \multicolumn{8}{c}{\textsc{CNNs}~(\textit{Few-shot})} \\
      \midrule
      \textsc{ResNet50}       & 150 & 77.40 & 32.80 & 13.50 & 18.30 & 29.58 & 37.80 \\
      \textsc{ResNext101}     & 150 & 83.07 & \textbf{38.80} & 16.80 & 21.30 & \textbf{47.96} & \textbf{56.25} \\
      \midrule
      \multicolumn{8}{c}{\textsc{InstructBLIP}~(\textit{Zero-shot})} \\
      \midrule
      BC                      & -  & 46.44 & 14.50 & 7.80 & 10.60 & 1.38 & 24.40 \\
      \midrule
      MC-TS (5)               & -  & 73.27 & 18.00 & 18.10 & 25.90 & 6.77 & 29.85 \\
      MC-TS (10)              & -  & 57.69 & 14.90 & 17.80 & 25.70 & 6.36 & 30.82 \\
      MC-TS (20)              & -  & -     & -     & 12.70 & 19.50 & 2.83 & 22.11 \\
      \midrule                
      OC                      & -  & 30.96 & 11.60 & 5.10  & 13.20 & 5.18 & 25.57 \\
      \midrule
      \multicolumn{8}{c}{\textsc{InstructBLIP}~(\textit{Few-shot})} \\
      \midrule
      BC                      & 150 & 67.31 & 16.40 & 15.50 & 17.80 & 6.70 & 18.80 \\
      \midrule
      MC-TS (5)               & 150 & \textbf{87.98} & 24.30 & 25.20 & 29.40 & 17.62 & 33.31 \\
      MC-TS (10)              & 150 & 87.12 & 23.90 & \textbf{26.60} & \textbf{30.70} & 17.00 & 33.10 \\
      MC-TS (20)              & 150 & -     & -     & 25.60 & 29.40 & 15.83 & 33.59 \\
      \midrule
      OC                      & 150 & 87.31 & 34.60 & 18.90 & 21.90 & 18.24 & 42.23 \\
      \bottomrule
    \end{tabularx}
  \end{table}

\begin{figure}[t]
    \imageTextRow{1.5cm}{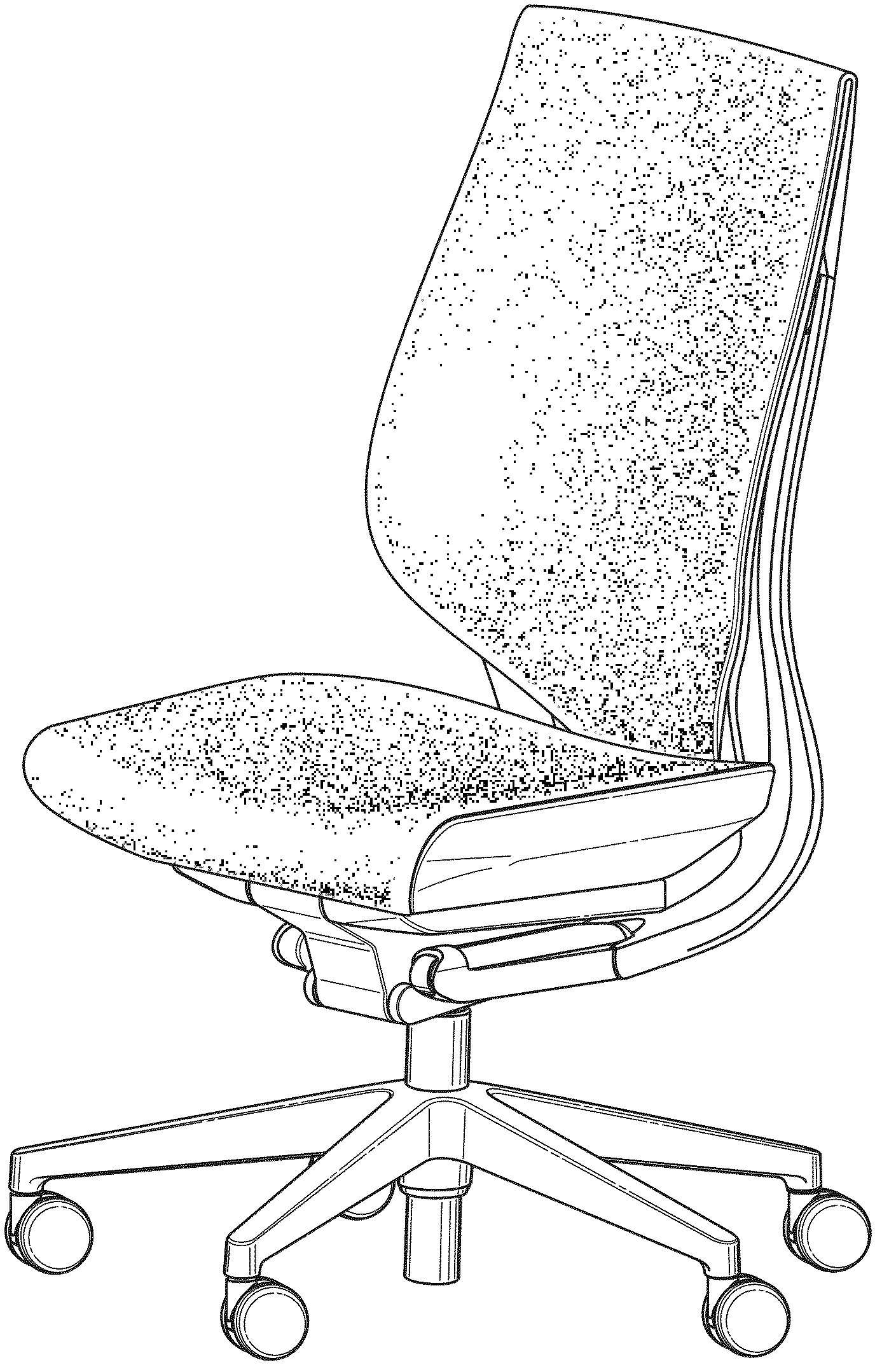}{
        \textbf{Ground Truth} $\rightarrow$ chair \newline
        \textbf{ResNext101} $\rightarrow$ office chair; Top-3 - [office chair, seating units, armchair] \newline
        \textbf{MC-TS (5)} $\rightarrow$ office chair; Last Level - [office chair, task chair, swivel chair]
    }
    \imageTextRow{1.5cm}{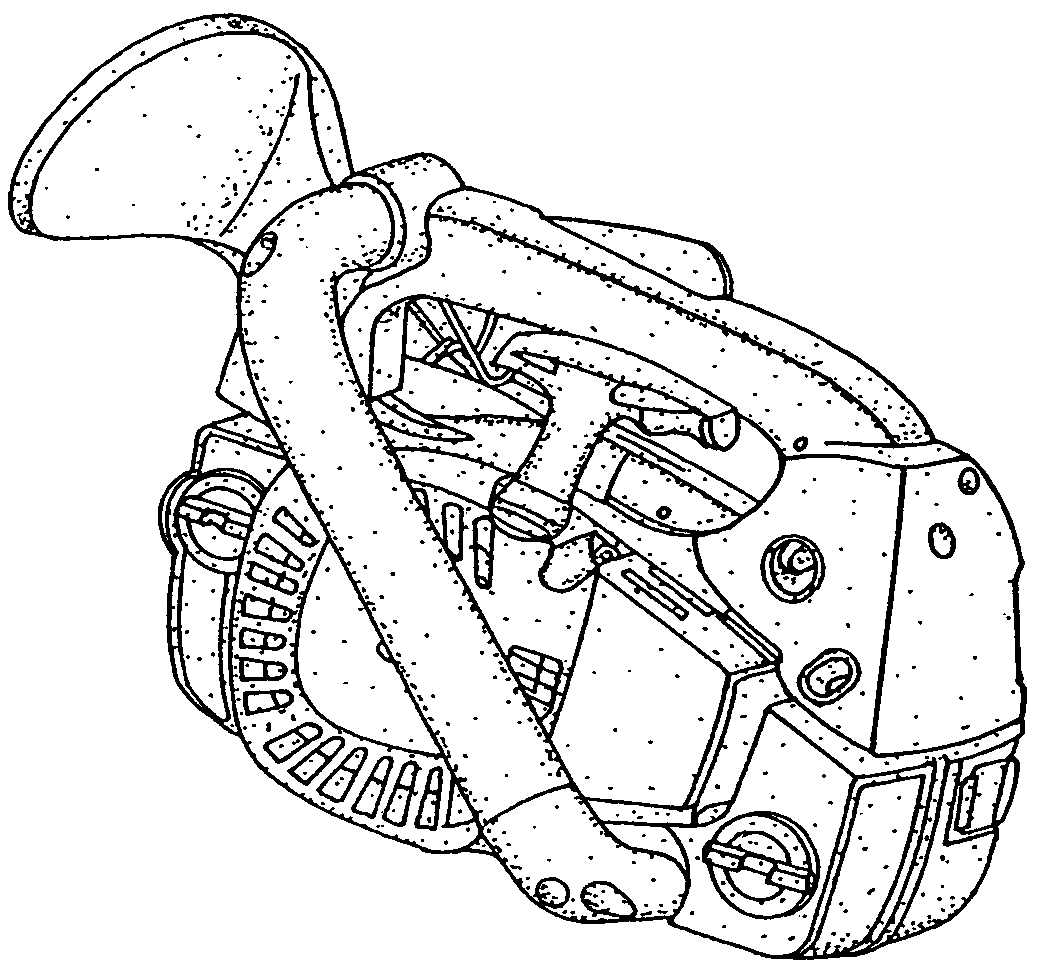}{
        \textbf{Ground Truth} $\rightarrow$ chainsaw \newline
        \textbf{ResNext101} $\rightarrow$ pet throw; Top-3 - [pet throw, robot, infant seat] \newline
        \textbf{MC-TS (5)} $\rightarrow$ chainsaw; Last Level - [electric bike, reciprocating saw, chainsaw]
    }
    \imageTextRow{1.5cm}{figures/samples/uspc_sample_1.png}{
        \textbf{Ground Truth} $\rightarrow$ edible products \newline
        \textbf{ResNext101} $\rightarrow$ equipment for preparing or serving food... ; Top-2 - [equipment for preparing or serving food..., cosmetic products and toilet articles] \newline
        \textbf{MC-TS (5)} $\rightarrow$ edible products; Last Level - [edible products, machines not elsewhere specified]
    }
    \imageTextRow{0.8cm}{figures/samples/uspc_sample_3.png}{
        \textbf{Ground Truth} $\rightarrow$ apparel and haberdashery \newline
        \textbf{ResNext101} $\rightarrow$ apparel and haberdashery; Top-2 - ["apparel and haberdashery", "equipment for safety, protection, and rescue"] \newline
        \textbf{MC-TS (5)} $\rightarrow$ games, toys, and sports goods; Last Level - ["games, toys, and sports goods", "furnishings"]
    }
    \caption{Qualitative examples for aspects~\textsc{Object} (rows 1 and 2) and~\textsc{USPC} (rows 3 and 4) comparing classification results of~\textsc{MC-TS} and~\textsc{ResNext101}}
    \label{fig:qualitative_analysis}
\end{figure}

Table~\ref{tab:cls_results} reveals that among the various LVLM-based classification methods, the \textsc{MC-TS} approach consistently outperforms the \textsc{BC} approach across all aspects. It only falls short of the \textsc{OC} approach in the \textsc{Projection} and \textsc{Object Seen} categories. Notably, \textsc{MC-TS} surpasses the CNN-based supervised baselines for \textsc{Type} and \textsc{USPC}.
\begin{figure*}[]
    \centering
    \begin{subfigure}{\textwidth}
        \subgraphscfmatrix{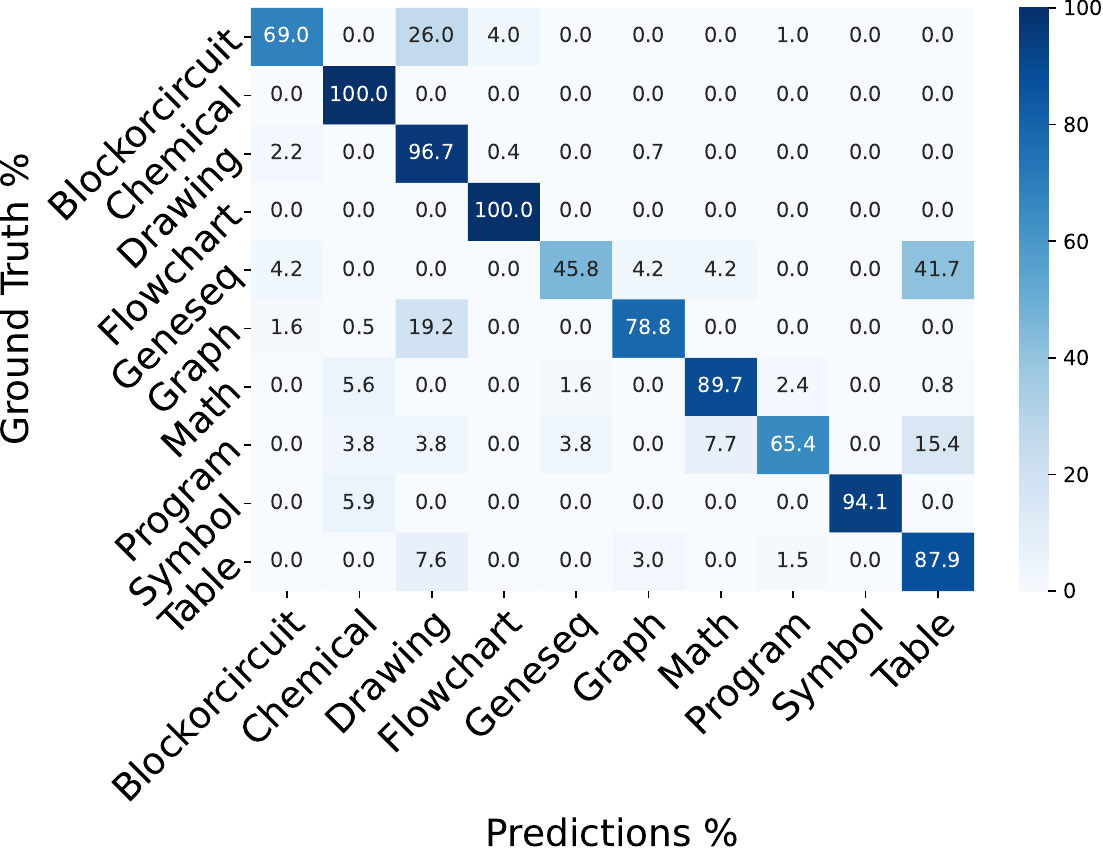}{}{fig:type_cm}
        \subgraphscfmatrix{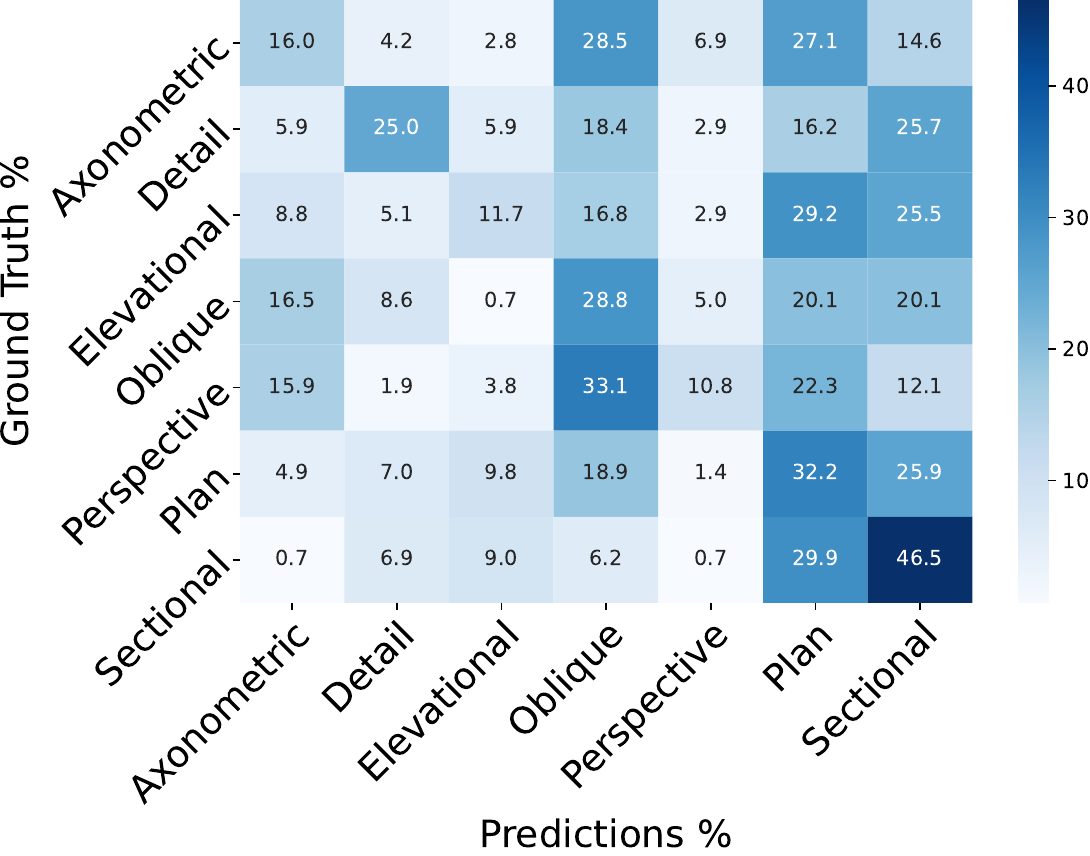}{}{fig:projection_cm}
    \end{subfigure}
    \caption{Confusion matrix on results produced by fine-tuned~\textsc{InstructBLIP} using~\textsc{MC-TS~(5)} approach for aspects~\textsc{Type} (left) and~\textsc{Projection} (right)}
    \label{fig:cf_type_projection}
\end{figure*}
While LVLM-based approaches generally lag behind CNN-based supervised models in terms of accuracy, it's important to note that this metric may not fully reflect the true performance. This is because the concepts for the figures were automatically extracted from figure references~\cite{Ajayi2023} without normalization. Figure~\ref{fig:qualitative_analysis} illustrates this point with randomly sampled figures for \textsc{Object} and \textsc{USPC} aspects, showing their corresponding ground truth and predicted concepts. We observe that although the selected concepts are often semantically relevant to the figure, they may not align perfectly with the ground truth.

To address this discrepancy, we introduced the \textsc{SemEq} metric. Using this metric, we observe a significant improvement in performance across all approaches, providing a more comprehensive evaluation of the models' capabilities in classifying patent figures.

For aspects~\textsc{Type} and~\textsc{USPC}, the~\textsc{MC-TS} approach outperforms the CNN-based classifier. In Figure~\ref{fig:cf_type_projection}, we observe how the LVLM is capable of classifying~\textsc{Type} and~\textsc{Projection} for patent figures. Similar to the performance in the VQA setting, the LVLM struggles to distinguish between the different projections in patent figures.

\section{Conclusion and Future Work}
\label{sec:conclusion_and_future_work}

In this paper, we have studied patent figure VQA and classification using LVLM in zero-shot and few-shot settings. We adapted an LVLM, primarily pre-trained on natural images, to the patent domain by fine-tuning it for VQA task in a few-shot learning scenario. For this purpose, we introduced a novel dataset --~\PatVQA{}-- suitable for fine-tuning and evaluation of an LVLM for patent figure VQA task. We have also introduced a novel LVLM-based tournament-style classification approach, that can handle large number of classification labels. The proposed solution outperforms or is comparable to other LVLM-based approaches, and even outperforms supervised CNN-classification models for aspects~\textsc{Type} and~\textsc{USPC} on a novel~\PatCLS{} test dataset. 

In future, it would be interesting to see how the proposed LVLM-based tournament-style classification approach fairs against the supervised CNN-based models with use of other LVLMs. Furthermore, fine-tuning more layers of an LVLM can improve results compared to CNNs that update all weights during training. Prompt compression could also be an interesting research direction to mitigate large context length for numerous classification labels.

\begin{credits}
\subsubsection{\ackname} This article has been funded by the Academic Research Programme of the European Patent Office (project "ViP@Scale: Visual and multimodal patent search at scale").
\end{credits}


%
%
\bibliographystyle{splncs04}
\bibliography{unified_bib.bib}

\end{document}